\documentclass[10pt,letterpaper]{article}
\usepackage{opex3}
\usepackage{cite}
\usepackage{hyperref}

\begin{document}

\title{Preventing and Reversing Vacuum-Induced Optical Losses in High-Finesse Tantalum (V) Oxide Mirror Coatings}

\author{Dorian Gangloff,$^{1,\dagger}$ Molu Shi,$^{1,\dagger}$ Tailin Wu,$^{1,\dagger}$ Alexei Bylinskii,$^{1}$ Boris Braverman,$^{1}$ Michael Gutierrez,$^{1}$ Rosanna Nichols,$^{1}$ Junru Li,$^{1}$ Kai Aichholz,$^{2}$ Marko Cetina,$^{1}$ Leon Karpa,$^{1}$ Branislav Jelenkovi\'{c}$^{3}$, Isaac Chuang$^{1,2}$, and Vladan Vuleti\'{c}$^{1,*}$}
\address{$^1$Department of Physics, MIT-Harvard Center for Ultracold Atoms and Research Laboratory of Electronics, Massachusetts Institute of Technology, Cambridge, Massachusetts 02139, USA \\ $^2$Department of Electrical Engineering and Computer Science, Massachusetts Institute of Technology, Cambridge, Massachusetts 02139, USA \\ $^3$Institute of Physics, University of Belgrade, Serbia \\ $^\dagger$These authors contributed equally to this work}
\email{*vuletic@mit.edu}

\date{\today}

\begin{abstract}
We study the vacuum-induced degradation of high-finesse optical cavities with mirror coatings composed of SiO$_2$-Ta$_{2}$O$_{5}$ dielectric stacks, and present methods to protect these coatings and to recover their initial quality factor. For separate coatings with reflectivities centered at 370~nm and 422~nm, a vacuum-induced continuous increase in optical loss occurs if the surface-layer coating is made of Ta$_{2}$O$_{5}$, while it does not occur if it is made of SiO$_2$. The incurred optical loss can be reversed by filling the vacuum chamber with oxygen at atmospheric pressure, and the recovery rate can be strongly accelerated by continuous laser illumination at 422~nm. Both the degradation and the recovery processes depend strongly on temperature. We find that a 1~nm-thick layer of SiO$_2$ passivating the Ta$_{2}$O$_{5}$ surface layer is sufficient to reduce the degradation rate by more than a factor of 10, strongly supporting surface oxygen depletion as the primary degradation mechanism.
\end{abstract}




\section{Introduction}
High-finesse mirrors are commonly used in a range of applications requiring high-vacuum environments. This includes ultrastable optical frequency references \cite{Kessler2012}, where spurious drifts in pressure, humidity and temperature are greatly reduced by placing the mirrors in vacuum, and atomic physics experiments \cite{Cetina2013, Sterk2012}, where the strong coupling of light to trapped atoms can be achieved using high-finesse optical cavities. The high mirror reflectivities required for these applications are predominantly achieved using dielectric stack structures of tantalum (V) oxide (Ta$_{2}$O$_{5}$) and silicon (IV) oxide (SiO$_{2}$) with layer spacings on the scale of the light wavelength \cite{Sites1983,Rempe1992}. The vacuum-facing layer is typically Ta$_{2}$O$_{5}$, owing to its higher index of refraction. It has been observed that, for this type of mirror, absorption losses increase dramatically over time under vacuum \cite{Cetina2013}, causing the finesse of these cavities to be reduced by a reported factor of 3 or more \cite{Cetina2011,Sterk2012}. When the mirror temperature is raised to $450^\circ$C, as required to anneal the mirrors under vacuum, measurements using light at infrared wavelengths indicate that the loss increase is accompanied by a reduction in the concentration of oxygen in the Ta$_{2}$O$_{5}$ surface layer of the mirror \cite{Brandstatter2013}. This observed oxygen depletion points towards a degradation process caused by changing levels of oxidation in the surface layer \cite{Brandstatter2013}, rather than impurity deposition. Although the temperature in the aforementioned study is higher than in most optical and atomic physics applications, optical losses resulting from oxygen depletion in Ta$_{2}$O$_{5}$ have been reported to be gradually more severe towards shorter wavelengths ($\leq 800$~nm) \cite{Demiryont1985}, for which the present study is conducted.

In this paper, we investigate the time-dependence of the vacuum-induced optical losses using an optical cavity formed by high-finesse mirrors placed under high-vacuum. We investigate the losses at multiple wavelengths (370~nm and 422~nm), for different temperatures (21$^\circ$C-150$^\circ$C) and different surface layers (Ta$_{2}$O$_{5}$ and SiO$_{2}$). We show that these losses can be partially or fully reversed by exposing the mirrors to a pure-oxygen environment, and introduce an oxygen-depletion model that is quantitatively supported by our observations. 

The structure of the paper is as follows. In Section 2, we introduce the experimental parameters and the measurement procedures. In Section 3, we observe that the rate of the loss increase is a steep function of temperature, and that this behavior is present both at 370~nm and 422~nm. In Section 4, we study the reversibility of the degradation process. We examine the cavity losses in chambers filled with pure oxygen, and find that the cavity finesse can be fully or partially recovered, indicating the crucial role played by surface oxygen in the loss process. In Section 5, we show that a photo-assisted process enhances this rate of recovery. In Section 6, we demonstrate the specificity of the degradation process to Ta$_{2}$O$_{5}$. We show that a 1~nm-thick layer of SiO$_{2}$ passivating the Ta$_{2}$O$_{5}$ top layer reduces the degradation rate by at least a factor of 10, while a 110~nm-thick layer prevents it altogether. Lastly, in Section 7, we present and discuss an oxygen-depletion model which is consistent with the literature and is in quantitative agreement with our data.


\section{Methods}

In this section, we describe the investigated coatings, the experimental apparatus and our measurement procedures.

\subsection{Mirror Coatings}

We perform experiments with four different coatings, designed for two wavelengths (370~nm and 422~nm) and employing two surface-layer materials (Ta$_{2}$O$_{5}$ and SiO$_{2}$). The first coating (Coating I-1) (deposited by Advanced Thin Films in Boulder, CO) has a Ta$_{2}$O$_{5}$ surface layer, and a reflectivity spectrum centered around the wavelength of 370~nm, where its transmission is 180~ppm. Coating I-2 consists of a 1~nm-thick layer of SiO$_{2}$ deposited on top of Coating I-1 (in-house deposition). Coating III-1 and Coating III-2 (deposited by Advanced Thin Films in Boulder, CO) have reflectivity spectra centered around 422~nm, where their transmissions are 40~ppm and 45~ppm respectively, and have surface-layers made of Ta$_{2}$O$_{5}$ and SiO$_{2}$ respectively. The surface layer thickness for each coating can be found in Table~\ref{table}.

\subsection{Experimental Setups for Two Wavelengths}

Three different experimental setups, each with high-finesse Fabry-Perot cavities constructed from mirrors with the described coatings, are used to measure the mirror losses: a vacuum chamber dedicated to testing 370~nm mirrors under vacuum (chamber I), an atomic physics setup with a 370~nm cavity under ultra-high vacuum \cite{Cetina2013} (chamber II), and a vacuum setup dedicated to simultaneously testing two pairs of 422~nm mirrors (chamber III). In each case, the two mirrors forming the cavity have the same coating. Schematics of the experimental setups for chamber I and chamber III can be found in Fig.~\ref{systemfig}a and Fig.~\ref{systemfig}b, respectively. 

In chamber I, mirrors with Coating I-1 and I-2 (tested separately) are placed under vacuum, at a pressure of $7\times10^{-6}$ Pa (maintained by an ion pump). The temperature of the mirrors is varied by heating the chamber and monitored using an external probe. In chamber II, mirrors with Coating I-1 are placed under vacuum, at a pressure of $10^{-8}$ Pa (maintained by an ion pump), and actively stabilized to 33$^\circ$C by measuring the temperature close to the mirror mount. In chamber III, two cavities with separate optical paths, consisting of mirrors with Coatings III-1 and III-2 respectively, are tested simultaneously. The chamber is heated and actively stabilized to 57$^\circ$C by measuring the temperature close to the mirror mount, and the pressure is maintained at approximately $7\times10^{-5}$ Pa with a turbo-molecular pump. 

A summary of the experimental parameters and the results presented in this paper can be found in Table \ref{table}.


\begin{table}[h]
\begin{center}
\caption{Summary of the experimental parameters. }
\begin{tabular}[b]{ | c ||| c | c | c || c | c | }
\hline
Wavelength & \multicolumn{3}{ c|| }{370~nm} & \multicolumn{2}{ c| }{422~nm}\\ \hline
Coating & I-1 & I-2 & I-1 & III-1 & III-2 \\ \hline
Top Layer & Ta$_2$O$_5$ & SiO$_2$ & Ta$_2$O$_5$ & Ta$_2$O$_5$ & SiO$_2$ \\ \hline
Top Layer Thickness (nm) & 28.3 & 1 & 28.3 & 48.6 & 110 \\ \hline
Transmission $\mathcal{T}$ (ppm) & 180 & 180 & 180 & 40 & 45 \\ \hline
Chamber & \multicolumn{2}{ c| }{I} & II & \multicolumn{2}{ c| }{III} \\ \hline
Pressure (Pa) & \multicolumn{2}{ c| }{$7\times10^{-6}$} & $10^{-8}$ & \multicolumn{2}{ c| }{$7\times10^{-5}$} \\ \hline
Cavity Length $\mathcal{L}$ (cm) & 5 & 5 & 2.2 & 4.1 & 2.2 \\ \hline
Temperature ($^\circ$C) & 21, 50, 75, 100, 150 & 100 & 33 & 57 & 57 \\ \hline
Loss Increase Observed & Yes & Yes (slow) & Yes & Yes & No \\ \hline
Fig. Reference & \ref{finesseVStemp}, \ref{timescaleVStemp}, \ref{O2recovery} & \ref{fig:422377SiTaCompFdecay} &  \ref{finesseVStemp}, \ref{timescaleVStemp} & \ref{422FdecayTa}, \ref{fig:422Photo}, \ref{fig:422377SiTaCompFdecay} & \ref{fig:422377SiTaCompFdecay} \\ \hline

\end{tabular}

\label{table}
\end{center} 
\end{table}

\subsection{Measuring Loss}

To determine the mirror loss for coatings described in Table I, we measure the time constant $\tau_c$ of the free-decay of the light intensity in the optical cavities. We use this information to calculate the loss of the mirrors, as described in Ref. \cite{Siegman1986}. In the absence of light input to the cavity, the intra-cavity light intensity $I$ as a function of $t$ follows an exponential decay: $I(t) = I_{0}\exp(-t/\tau_{c})$. Here, $\tau_{c}$ is the cavity decay time, and $I_0$ is the intra-cavity intensity at the time light input is turned off. Given the length of the cavity $L$, the speed of light in vacuum $c$, and the mirror transmission factor $\mathcal{T}$, the loss $\mathcal{L}$ per mirror is given by $\mathcal{L} = L/(c\tau_{c})-\mathcal{T}$. This assumes identical loss and transmission for the two mirrors forming the cavity. Although this is an approximation which assumes uniformity of manufacturing, this assumption does not affect the time- and temperature-dependence of the measured loss increase.

\begin{figure}[h!!]
\centering\includegraphics[width=12cm]{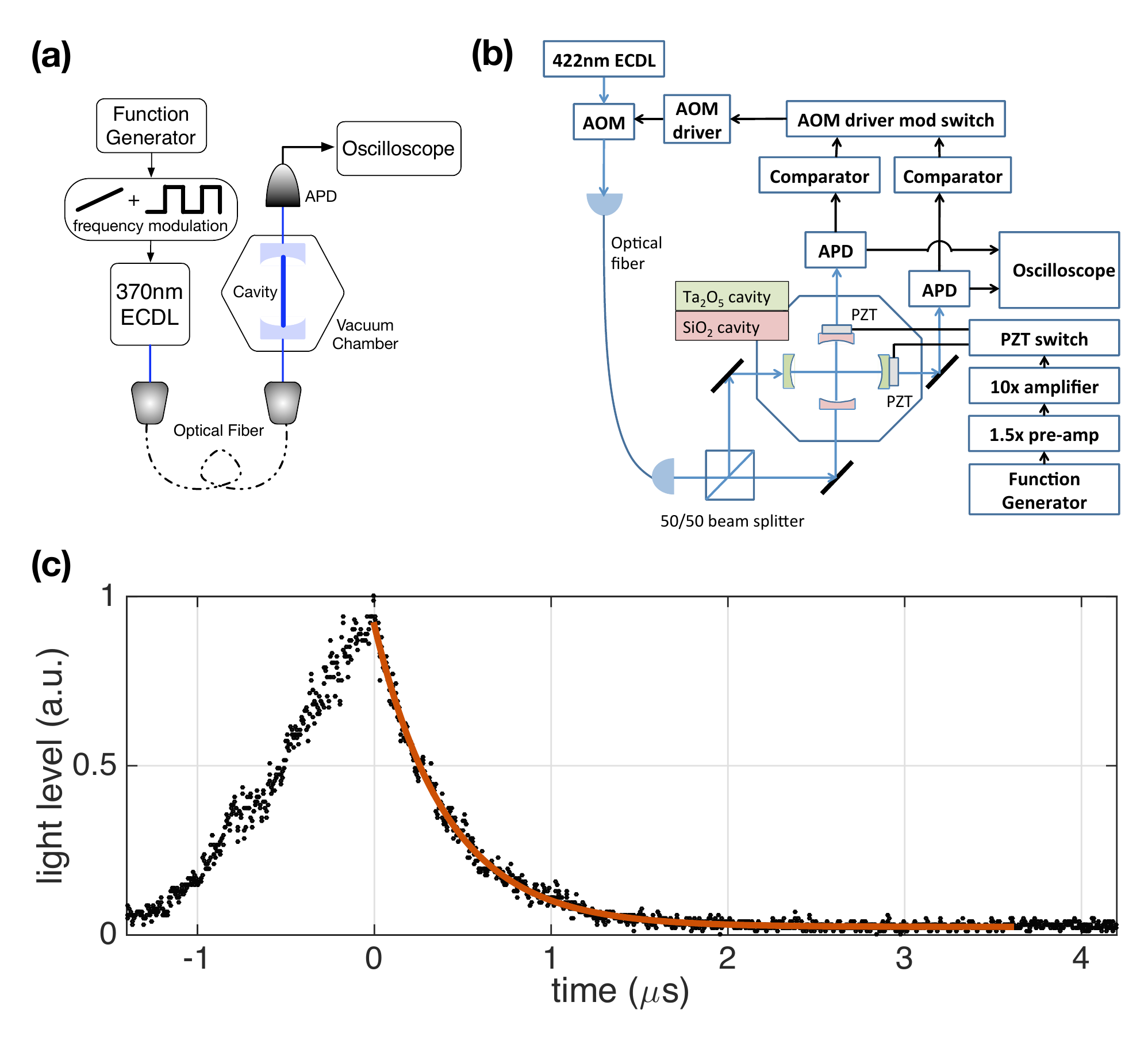}
\caption{(a) Schematic of the experimental setup for chamber I. A pair of mirrors forming a high-finesse cavity with either Ta$_{2}$O$_{5}$ (Coating I-1) or SiO$_2$ (Coating I-2) as their surface layer (tested separately), are placed under high vacuum. Light from a single-mode laser at 370~nm is used to probe these cavities as the laser frequency is slowly and linearly scanned by a function generator. The transmitted light from the cavity is incident on an avalanche photodiode (APD). The laser frequency is also modulated by a fast square-wave signal, which results in a free-decay of the cavity's transmitted light intensity each time the slow scan brings the laser in resonance with the cavity (time = 0~$\mu$s). (b) Schematic of the experimental setup for chamber III. Two pairs of mirrors forming high finesse cavities with SiO$_2$ (Coating III-2) and Ta$_{2}$O$_{5}$ (Coating III-1) as their surface layer, respectively, are placed under high vacuum and tested simultaneously. Light from a single mode laser at 422~nm is used to probe the cavities as they are scanned using piezoelectric transducers (PZT). The transmitted light from each cavity is incident on an avalanche photodiode (APD). When the cavity becomes resonant with the laser, and the signal intensity reaches a defined threshold in a comparator, the laser light is switched to be off-resonant using an accousto-optic modulator (AOM) (time = 0~$\mu$s), resulting in a free-decay of the cavity's transmitted light intensity. (c) A typical light intensity free-decay curve measured for 370~nm (Coating I-1), fitted with an exponential model with a time constant of $\tau_{c} = 411$~ns.
\label{systemfig}}
\end{figure}

Free-decay traces are obtained by driving the cavity with resonant laser light, switching the light input off much faster than $\tau_{c}$, and observing the relaxation of the light intensity at the cavity output. We use extended-cavity laser diodes (ECDLs) as narrow-band single mode sources of light whose linewidths ($\sim 2.5$ MHz \cite{Loh2006}) are comparable to those of the cavities investigated, allowing for efficient excitation of the cavity and high signal-to-noise ratio of the free-decay intensity $I(t)$. The transmitted intensity is measured using avalanche photodiodes with sufficient bandwidths ($\sim$20~MHz) to capture signals changing much faster than $\tau_{c}$. The switching of the input light is done in different ways for cavities in chambers I and II, and in chamber III, respectively. For cavities in chambers I (see Fig.~\ref{systemfig}a) and II \cite{Cetina2013}, the frequency of the ECDL is scanned across the modes of the cavity (the ECDL and the cavity resonant frequencies are not actively stabilized to one another). Simultaneously, a square-wave current modulation is applied to the laser diode, causing it to switch frequency by an amount much larger than the linewidth of the cavity, at a rate that is higher than the scan rate but smaller than the cavity linewidth. When the laser is scanned over the cavity resonance, the rise in intracavity intensity is interrupted by the laser's frequency switching, resulting in a free decay of the light transmitted through the cavity. This decay is detected using an oscilloscope triggered on the negative slope of the photodiode signal. In experiments performed in chamber III (see Fig.~\ref{systemfig}b), the cavity length, and hence the cavity resonant frequency, is scanned with a piezoelectric transducer (PZT) mounted at the back of a cavity mirror (the ECDL and the cavity resonant frequencies are not actively stabilized to one another). Laser light incident onto the cavity is turned off with an accousto-optic modulator once a cavity mode becomes resonant with the frequency of the incident laser light; this occurs when the transmitted light intensity reaches a predefined threshold. Because spurious triggering events can occur, in both setups, only decays which reach their maximum intensity value at the edge from the pulse drive are considered. For these decays, we fit the dependence of the transmitted light intensity on time with an exponential model and extract the time constant of the intensity free-decay $\tau_{c}$. A sample free-decay curve is shown in Fig.~\ref{systemfig}c. 

\begin{figure}[h!!]
\centering\includegraphics[width=11cm]{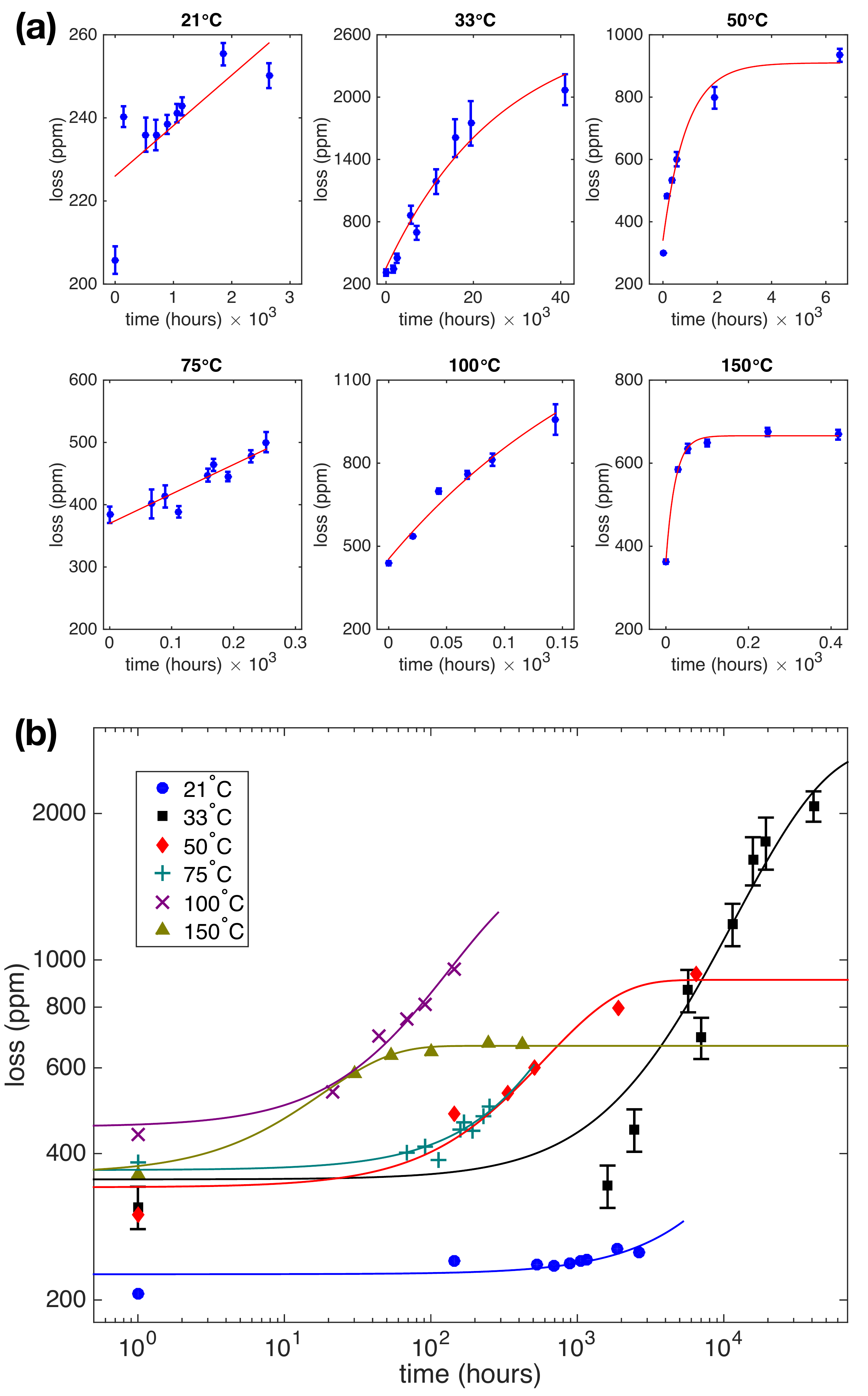}
\caption{Increase of loss of mirrors with Coating I-1 over time at various temperatures $T$, as separate panels on a linear scale (a) and combined on a log-log scale (b): $T = $21$^{\circ}$C, 50$^{\circ}$C, 75$^{\circ}$C, 100$^{\circ}$C, 150$^{\circ}$C (chamber I); and 33$^{\circ}$C (chamber II). Each data set is fitted with an exponential model shown as a solid line (a,b). Error bars are statistical and correspond to one standard deviation (smaller than the size of the data symbol when not shown).
\label{finesseVStemp}}
\end{figure}

The last four loss values measured for Coating I-1 at 33$^{\circ}$C (Fig. \ref{finesseVStemp}a) are obtained by measuring the cavity finesse $\mathcal{F} = \nu_{FSR}/(\kappa/2\pi)$, where $\nu_{FSR} = c/2L$ is the cavity's free spectral range, and $\kappa = 1/\tau_c$ is its linewidth \cite{Siegman1986}. The linewidth $\kappa > 1$~MHz and the free spectral range are measured simultaneously by linearly scanning a frequency-doubled titanium-sapphire laser, whose linewidth is less than 100~kHz, across the cavity resonances. The finesse can be used to determine the loss as $\mathcal{L} = \pi/\mathcal{F} - \mathcal{T}$.

\section{Loss increase}

In this section, we present our results on the time-dependence of the vacuum-induced losses, and investigate the rate of the increase in losses as a function of temperature and light wavelength.

\subsection{Temperature Dependence}

Figure~\ref{finesseVStemp} shows the mirror losses as a function of time for different temperatures $T$. The temperature of the mirrors is varied from 21$^{\circ}$C to 150$^{\circ}$C in chamber I, and kept at 33$^{\circ}$C in chamber II. The observation times range from a few days ($T=$100$^{\circ}$C) to a few years ($T=$33$^{\circ}$C). In all experiments, we observe an increase of loss with time. While the loss initially increases linearly in time, for data sets taken over sufficiently long times, we observe that the loss saturates. We find that the typical time scale for the loss to increase, and to reach saturation, sharply decreases with temperature. At 21$^{\circ}$C, the loss increases by only 20\% after 12 weeks, while at 150$^{\circ}$C, the loss saturates at about twice its initial value after just 3 days.

In Fig.~\ref{finesseVStemp}, we also show the fit of the time dependence of the loss increase to an exponential model with three free parameters: $\mathcal{L}(t) = \mathcal{L}(0) + \Delta \mathcal{L} (1 - \exp(-t/\tau_{th}))$ (see Section 7), where $\tau_{th}$ is the time scale of the loss increase.

\begin{figure}[h!]
\centering\includegraphics[width=10cm]{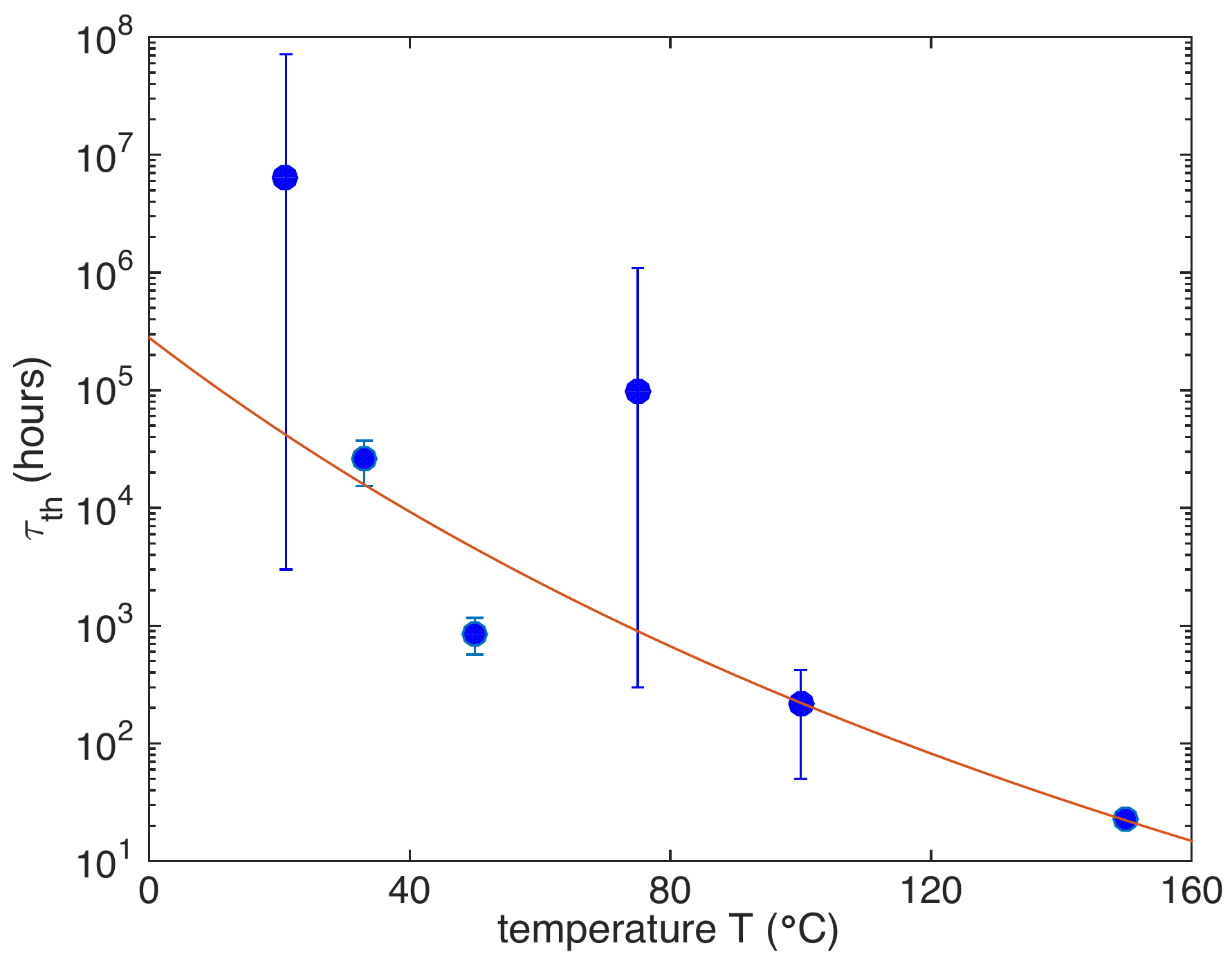}
\caption{(Coating I-1) Time scale of the loss increase $\tau_{th}$, for the data from Fig.~\ref{finesseVStemp}, depending on temperature $T$. We fit the data with a model of the form $\tau_{th} = \tau_0 \exp(a/(273+T))$ (red solid line); the fitted values are $a = 7300(1600)$~K and $\ln(\tau_0)=-14(4)$. The fit is weighted by the inverse error variance on each data point. Parentheses and error bars indicate a 68\% confidence interval on the fitted values.
\label{timescaleVStemp}}
\end{figure}

Figure~\ref{timescaleVStemp} shows the dependence of this time scale on temperature $T$, and a fit to a model where $\tau_{th}$ depends exponentially on the inverse of $T$. This relationship is highly suggestive of an Arrhenius-type thermal activation of the process causing degradation, for which the thermal activation rate is $1/\tau_{th} \propto \exp(-U/k_BT)$, where $U$ is the activation energy and $k_B$ the Boltzmann constant. We find the activation temperature $U/k_B = 7300(1600)$~K, corresponding to an activation energy of $U=0.6(1)$~eV (68\% confidence interval). This model is consistent with oxygen depletion being the cause of the observed increase in loss, as discussed in Section 7.

We note that the loss saturation level $\mathcal{L}(0) + \Delta \mathcal{L}$ also appears to depend inversely on temperature. The data sets for which a saturation is observed (see Fig.~\ref{finesseVStemp}: 33$^{\circ}$C, 50$^{\circ}$C, and 150$^{\circ}$C) suggest that higher temperatures lead to lower loss saturation levels. This may be a result of a temperature-induced shift in the light absorption spectrum of the color centers that are responsible for the loss increase. This could be verified in a future experiment in which the vacuum-induced loss increase is measured at a high temperature until saturation of the loss is reached, followed by rapidly lowering the mirror temperature while measuring loss.

\subsection{Wavelength Dependence}

\begin{figure}[]
\centering\includegraphics[width=10cm]{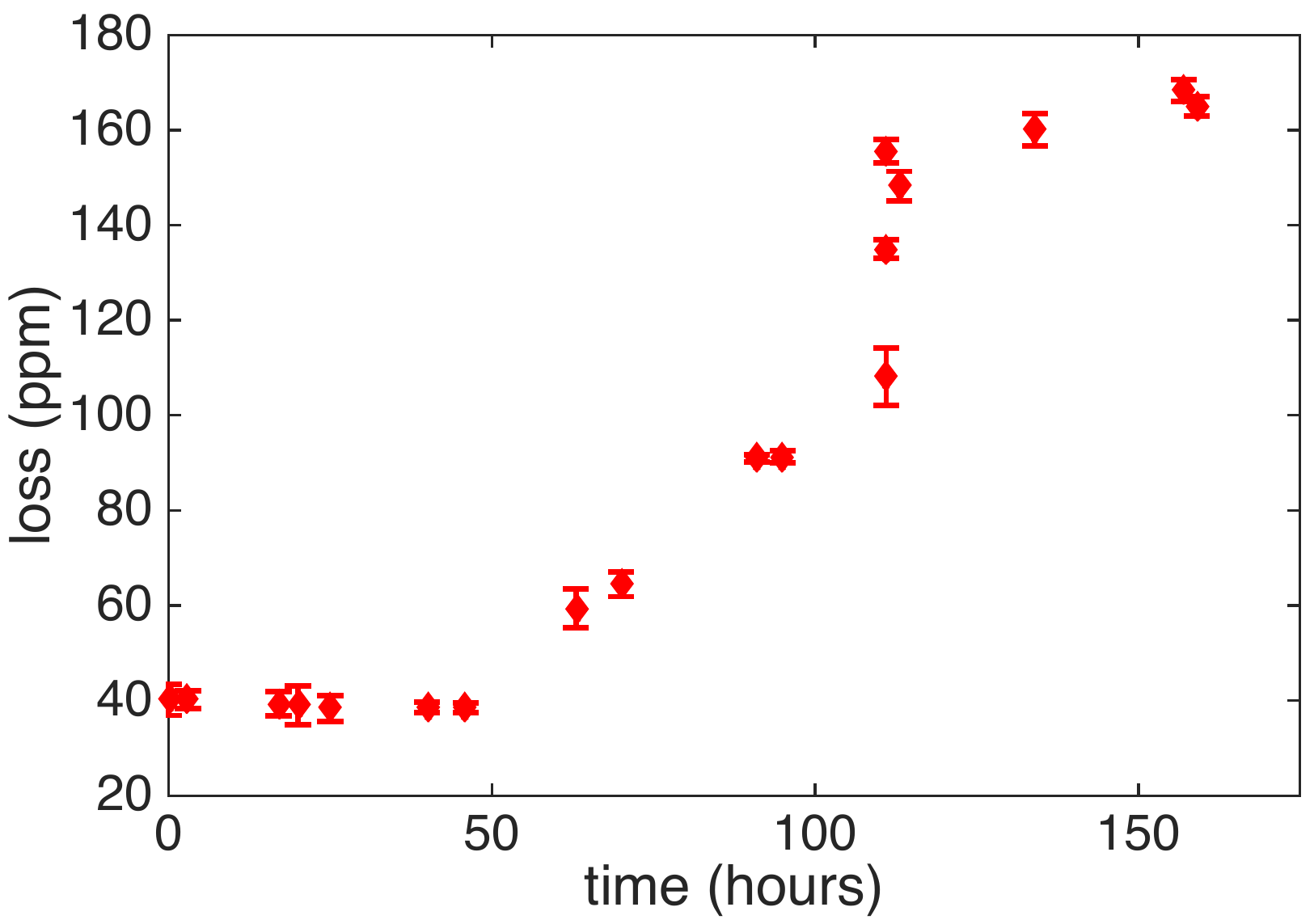}
\caption{Increase of optical loss observed for an optical cavity composed of mirrors with Coating III-1 (422~nm, Ta$_{2}$O$_{5}$ surface layer) at 57$^{\circ}$C in chamber III. Error bars are statistical and correspond to one standard deviation.}
\label{422FdecayTa}
\end{figure}

In Fig.~ \ref{422FdecayTa}, we investigate the increase of loss of mirrors employing Coating III-1 using light at 422~nm (chamber III). The measured time scale of the loss increase at 57$^{\circ}$C is much shorter than that measured using 370~nm light on Coating I-1 at comparable temperatures. This indicates a dependence of the rate of loss increase on wavelength, as well as temperature.

With the current data, we do not have a good explanation for the initial slow loss increase, which is observed only in this data set. It could be attributed to the thermal relaxation time of the mirror in chamber III.

\section{Recovery from Losses Using Oxygen}

In this section, we demonstrate that the presence of oxygen gas at the mirror surface can reverse the losses measured in the previous section. 

Reversal of the vacuum-induced loss is achieved by leaking high purity oxygen (Airgas Ultrahigh Purity Grade 4.4) into the test chamber via a needle valve, and monitoring loss for various temperatures and partial pressures of oxygen. 

\begin{figure}[h!!]
\centering
\includegraphics[width=12cm]{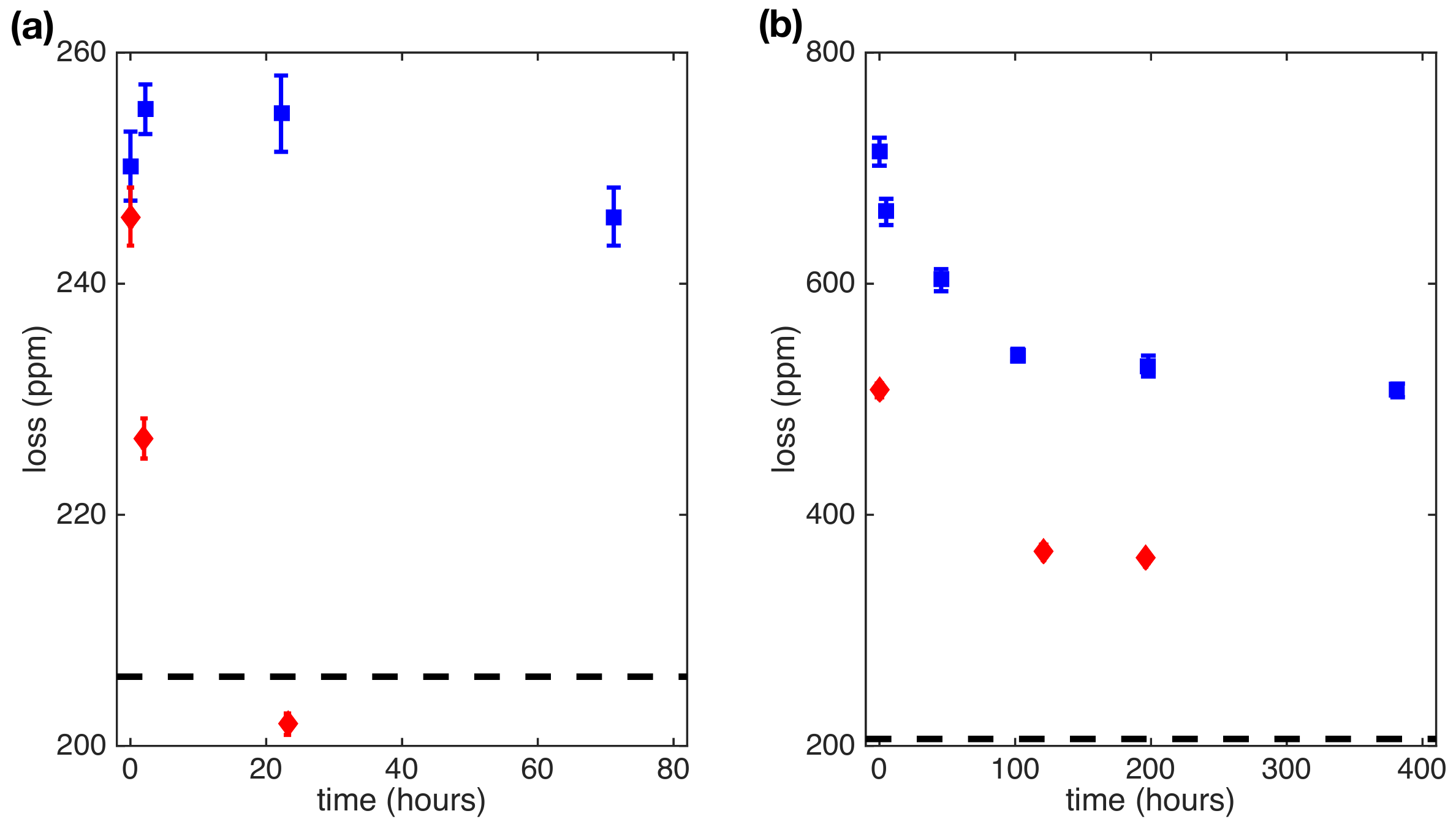}
\caption{(Coating I-1) (a) Recovery from vacuum-induced losses with oxygen, while at a temperature of 21$^{\circ}$C, following the data set at 21$^{\circ}$C (Fig.~\ref{finesseVStemp}). Shown are the loss under oxygen at a partial pressure of $10^{-2}$ Pa (blue squares), and loss under an atmospheric pressure of oxygen (red diamonds). (b) Recovery with an atmospheric pressure of oxygen, while at a temperature of 21$^{\circ}$C (blue squares) and 150$^{\circ}$C (red diamonds), following a vacuum-induced loss increase at a much higher temperature of 150$^{\circ}$C (data not shown). The dashed lines indicate the loss value prior to oxygen treatment. Error bars are statistical and correspond to one standard deviation.
\label{O2recovery}}
\end{figure}

The blue squares on Fig.~\ref{O2recovery}a represent the dependence of the losses of mirrors with Coating I-1 during exposure to a partial pressure of oxygen of $10^{-2}$~Pa. The data are taken directly following the observation of vacuum-induced losses at 21$^{\circ}$C (data shown in Fig.~\ref{finesseVStemp}). Here, we observe a slight recovery from vacuum-induced losses. This is to be compared with the red diamonds on Fig.~\ref{O2recovery}a, which represent the losses during exposure to an atmospheric pressure of oxygen. There, we observe a full recovery, taking approximately 10~hours, to the loss value of $\sim 205$~ppm measured before putting the mirrors under vacuum (dashed line in Fig.~\ref{O2recovery}a).

Following the observation of vacuum-induced losses in Coating I-1 up to $\sim 700$~ppm at a much higher temperature (approximately 150$^{\circ}$C), we repeat the recovery using oxygen. The blue squares on Fig.~\ref{O2recovery}b represent the dependence of the losses during exposure to an atmospheric pressure of oxygen while the mirrors are at a temperature of 21$^{\circ}$C. Here, we observe a recovery that takes $\sim 100$ hours, which is an order of magnitude slower compared to the recovery in Fig.~\ref{O2recovery}a, and only to a value of $\sim 500$~ppm. The red diamonds on Fig.~\ref{O2recovery}b represent the losses during exposure to an atmospheric pressure of oxygen while the mirrors are at a temperature of 150$^{\circ}$C, observed directly following the observation of losses shown as the blue squares in Fig.~\ref{O2recovery}b. There the loss returns to a value of $\sim 350$~ppm with a time constant of approximately $\sim 100$~hours, which still represents a partial recovery compared to the loss value of $\sim 205$~ppm measured before putting the mirrors under vacuum (dashed line in Fig.~\ref{O2recovery}b) . 

In summary, we observe partial or full recovery from the vacuum-induced loss for all cases tested. This is further evidence that oxygen concentration at the mirror surface is a determining component of the degradation process. It is unclear from the data presented in Fig.~\ref{O2recovery}b whether the loss increase at high temperatures is activated by additional processes, such as the action of surface contaminants, which prevent the full recovery with ambient oxygen gas, or whether the deeper oxygen depletion at high temperatures (such as in Fig.~\ref{finesseVStemp}, 150$^{\circ}$C) creates an additional energy barrier to oxygen re-entering the top layer of the coating (i.e. the oxygen binding process could be hysteretic).

\section{Photo-assisted Recovery Process}

In this section, we show that continuous illumination of the mirrors with near-UV light can dramatically accelerate the recovery rate under oxygen.

\begin{figure}[h!]
\includegraphics[width=12.5cm]{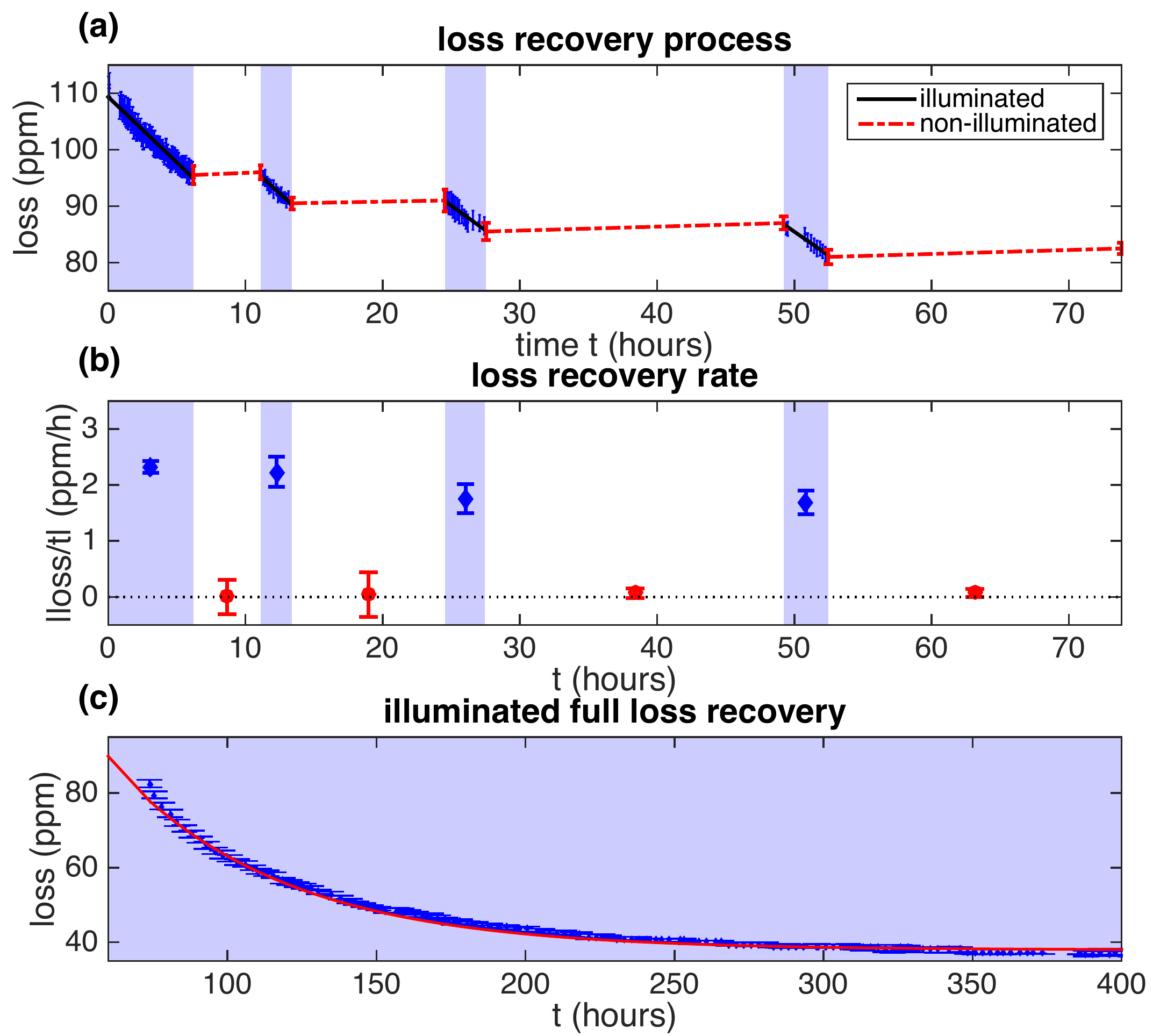}
\caption{Laser-assisted loss recovery processes observed for Coating III-1. (a) loss recovery observed during both illuminated and non-illuminated periods. (b) recovery rate obtained by fitting data in (a) using a linear model, for both the illuminated and non-illuminated periods. (c) optical loss is fully reversed by continuous illumination with an exponential time constant of 56.5~hours. Error bars are statistical and correspond to one standard deviation.}
\label{fig:422Photo}
\end{figure}

The reversal of vacuum-induced losses by the presence of oxygen suggests that a re-oxidation process of the mirror surface oxide might be involved. Studies of the dielectric thin film growth process indicate that the oxidation rate can be significantly affected by the presence of UV light illumination \cite{Zhang1998, Boyd2001}, especially for Ta$_2$O$_5$, for which improvement of the optical properties was found during a UV annealing stage. It was observed that under 172~nm radiation, oxygen can be easily dissociated to form stronger oxidizers, such as ozone or single O atoms, which can further oxidize defects, e.g.~suboxides of Ta, and increase the material transparency. In light of this, we examined whether recovery from vacuum-induced losses can be affected by illuminating the cavity with a resonant laser at 422~nm (the near-UV range). 

In chamber III, under atmospheric pressure of oxygen and at a temperature of 57$^{\circ}$C, we investigated the recovery from vacuum-induced losses (shown in Fig.~\ref{422FdecayTa}) for two controlled processes, as shown in Fig.~\ref{fig:422Photo}a: 1) an illuminated process (shown as a shaded area), where the cavity was continuously illuminated by a probe laser at 422~nm with about 10~kW/cm$^{2}$ of intra-cavity intensity; and 2) a non-illuminated process, where the same cavity was illuminated only at the beginning and end of a given time interval. We alternated the two processes four times. Figure~\ref{fig:422Photo}b shows the two corresponding recovery rates to be significantly different. The blue diamonds show a 2-3 ppm/hr recovery rate found during the illuminated periods, while the red circles show a negligible rate found during the non-illuminated periods. Figure~\ref{fig:422Photo}c shows a subsequent recovery process under constant illumination, resulting in a full recovery of the initial cavity loss level of $\sim 40$~ppm (see Fig.~\ref{422FdecayTa}).

The negligible rate of recovery observed here in the absence of laser light behaves like the slow rate of recovery observed following the 150$^{\circ}$C loss increase at 370~nm (Fig.~\ref{O2recovery}b). In both cases, the more than two-fold increase in the loss factor when under vacuum could be accompanied by an additional process with an energy barrier that prevents or slows down the re-entry of oxygen into the surface layer of the coating. This is a possible explanation for why UV light at 422~nm dramatically enhanced the rate of recovery, while a higher temperature enhanced the recovery level at 370~nm (see Section 7).

\section{Dependence of Loss on the Surface Material and Passivation with SiO$_{2}$}

In this section, we show that the loss increase is specific to a Ta$_2$O$_5$ surface layer, and that passivation with SiO$_2$ can strongly reduce the increase in loss observed in Section 3.

\begin{figure}[h!]
\centering
\includegraphics[width=13cm]{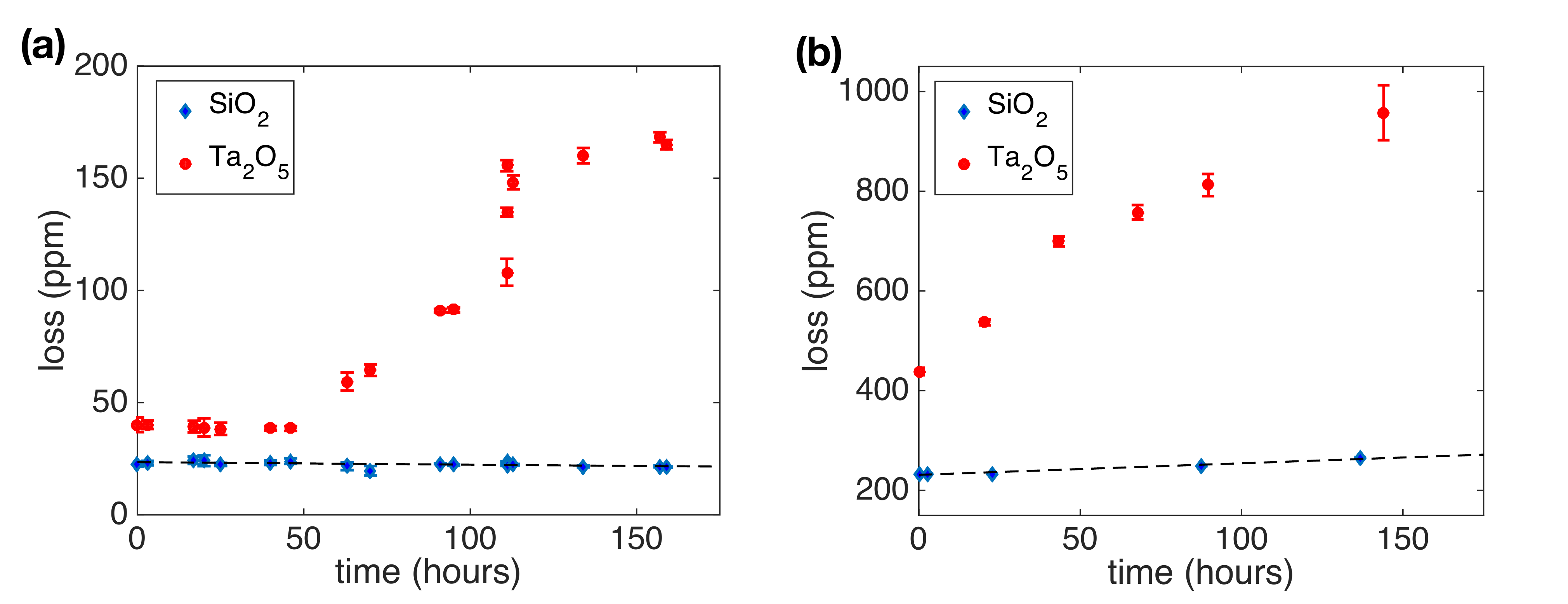}
\caption{The dependence of loss on time for different mirror top layers: (a) Loss increase at 57$^{\circ}$C measured at 422~nm for a Ta$_{2}$O$_{5}$ top layer (Coating III-1, red circles, data also shown in Fig.~\ref{422FdecayTa}), and for a 110~nm-thick SiO$_2$ top layer (Coating III-2, blue diamonds). The dashed line is a linear fit with a slope of $-0.011(4)$~ppm/h. (b) Loss at 100$^{\circ}$C measured at 370~nm for a Ta$_{2}$O$_{5}$ top layer (Coating I-1, red circles), and for a 1~nm-thick SiO$_2$ top layer (Coating I-2, blue diamonds). The dashed line is a linear fit with a slope of $0.23(3)$~ppm/h. Error bars are statistical and correspond to one standard deviation.}
\label{fig:422377SiTaCompFdecay}
\end{figure}

Figure~\ref{fig:422377SiTaCompFdecay}a shows the loss increase under vacuum (in chamber III) for mirrors with a Ta$_{2}$O$_{5}$ top layer (Coating III-1, red circles) and for mirrors with a 110~nm-thick SiO$_{2}$ top layer (Coating III-2, blue diamonds). At the same temperature (57$^{\circ}$C) and pressure ($7\times10^{-5}$~Pa), the mirrors with a Ta$_{2}$O$_{5}$ top layer show a significant loss increase, whereas the mirrors with a 110~nm-thick SiO$_{2}$ top layer show no loss increase. The dashed line of Fig.~\ref{fig:422377SiTaCompFdecay}a is a linear fit of the Coating III-2 data with a slope of $-0.011(4)$~ppm/h. This should be compared to the average loss increase of $\sim 1$~ppm/h for Coating III-1. These results demonstrate that the mirror coating degradation processes are strongly dependent on the surface layer material. Since SiO$_2$ has a higher activation energy for oxygen vacancy formation \cite{interactiveEllingham,Kubaschewski1967}, this is a further indication that surface oxygen plays a key role in the loss increase (see Section 7). 

The observed dependence of the loss increase on the surface material implies that passivating the Ta$_{2}$O$_{5}$ mirror coating with SiO$_{2}$ can prevent the increase of optical loss in vacuum. To test this idea, we sputter a thin layer of SiO$_{2}$ onto two mirrors with Coating I-1, resulting in Coating I-2. Based on the calibration of the sputtering machine, we estimate the thickness of the sputtered SiO$_2$ layer to be 1~nm. Figure~\ref{fig:422377SiTaCompFdecay}b shows the loss increase under vacuum (in chamber I) for mirrors with a Ta$_{2}$O$_{5}$ top layer (Coating I-1, red circles), and for the processed mirrors with a 1~nm-thick SiO$_{2}$ top layer (Coating I-2, blue diamonds). At the same temperature (100$^{\circ}$C) and pressure, the mirrors with a Ta$_{2}$O$_{5}$ top layer show a significant rate of loss increase, while the mirrors with a 1~nm-thick SiO$_{2}$ top layer have a much reduced rate of loss increase. The dashed line of Fig.~\ref{fig:422377SiTaCompFdecay}b is a linear fit of the Coating I-2 data. The resulting slope of $0.23(3)$~ppm/h should be compared to the average loss increase of $\sim 4$~ppm/h for Coating I-1.

While the data in Fig.~\ref{fig:422377SiTaCompFdecay}a and Fig.~\ref{fig:422377SiTaCompFdecay}b, respectively, are taken at different wavelengths and temperatures, the data for the mirrors with a Ta$_{2}$O$_{5}$ top layer (red circles) exhibit a similar average loss increase rate. This should be compared with the data for the mirrors with a SiO$_2$ top layer (blue diamonds), where a 1~nm-thick SiO$_2$ layer (Fig.~\ref{fig:422377SiTaCompFdecay}b) exhibits a measurable loss increase rate, while a 110~nm-thick SiO$_2$ layer (Fig.~\ref{fig:422377SiTaCompFdecay}a) exhibits no loss increase (within the measurement errors). This shows that the vacuum-induced loss can be completely suppressed by a sufficiently thick SiO$_2$ top layer. This dependence of the loss increase rate on the thickness of the surface SiO$_2$ layer also confirms that the loss process is due to a material transformation in the Ta$_{2}$O$_{5}$ surface layer affecting its optical properties, rather than due to an unknown process depositing absorbents onto the mirror surfaces. This observation is the strongest evidence we have for oxygen-depletion causing additional losses in the Ta$_{2}$O$_{5}$ surface layer.

\section{Model \& Discussion}

\subsection{Oxygen-depletion Model}

In this section we consider a model in quantitative agreement with our results that can explain the observed increase in optical losses, the subsequent recovery with oxygen treatment, and the dependence of both on temperature, incident light, and surface layer material. 

Oxygen vacancies at the dielectric stack surface can form as a result of an oxygen reduction process, creating color centers that increase the absorption losses in the mirror \cite{Demiryont1985}. At the surface, the oxygen is bound as an oxide which can form either free radicals or water via a redox reaction mediated by hydrogen ions. These reaction products quickly diffuse into vacuum making the reverse process highly improbable. This Arrhenius-type process is thermally activated, as seen in Fig.~\ref{timescaleVStemp}, and this process is reversible when vacuum is replaced by a sufficiently large partial pressure of oxygen, as seen in Fig.~\ref{O2recovery}. The presence of blue light can catalyze the formation of oxides \cite{Zhang1998, Boyd2001}, thus accelerating the reverse oxidation during the recovery process with ambient oxygen, as seen in Fig.~\ref{fig:422Photo}. The likelihood of the oxygen reduction process also strongly depends on the oxide making up the vacuum-facing surface layer, as seen in Fig.~\ref{fig:422377SiTaCompFdecay}. This agrees with the observation that the Gibbs free energy difference $\Delta G$ for the formation of SiO$_{2}$ is larger than for Ta$_{2}$O$_{5}$ \cite{interactiveEllingham,Kubaschewski1967}. At the temperature of $300$~K the oxidation Si $\rightarrow$ SiO$_2$ has $\Delta G \sim -850$~kJ/mol (or $-8.8$~eV) with a single intermediate oxide, while the oxidation Ta $\rightarrow$ Ta$_{2}$O$_{5}$ has $\Delta G \sim -750$~kJ/mol (or $-7.8$~eV) with four times the number of intermediate oxides.

The following gives further support to this model. 
Suboxide films of Ta$_{2}$O$_{5}$ were found to be absorbing and dispersive, with a strong dependence on oxygen content, in comparison to its non-absorbing stoichiometric counterpart \cite{Demiryont1985}. 
This is corroborated by a recent study \cite{Brandstatter2013} of high finesse IR mirrors, with Ta$_{2}$O$_{5}$ as their surface layer, placed under vacuum. 
Their surface concentration of oxygen, as measured by X-ray photoelectric spectroscopy (XPS), was found to decrease in high vacuum as the optical scattering losses increased.

Based on the above observations, we can construct a quantitative model for the loss increase over time as caused by oxygen depletion from the Ta$_{2}$O$_{5}$ surface layer. Given an incoherent light absorption process in the oxygen vacancy centers of the Ta$_{2}$O$_{5}$ film, the vacuum-induced absorption loss $\mathcal{L}$ of the tested mirrors would depend linearly on the oxygen vacancy concentration $\theta$ (as long as $\mathcal{L}\ll 1$): 
$\mathcal{L} = \mathcal{L}_0 + \mathcal{L}_1\theta(t)$.
For an Arrhenius process at a fixed temperature, the concentration of oxygen vacancies would follow an exponential in time, $\theta(t) \propto 1- \exp(-t/\tau)$. Combining this expression with our linear model of mirror loss, and absorbing the constants, we obtain 
$\mathcal{L}(t) = \mathcal{L}(0) + \Delta \mathcal{L} (1 - \exp(-t/\tau_{th}))$,
which has three free parameters and is used to fit our data (Fig.~\ref{finesseVStemp}).
Such a model is consistent with a linear increase in loss at small times compared to the typical time scale $\tau_{th}$, and a saturation to a finite value as time grows large, which we observe in our data (Fig.~\ref{finesseVStemp}). 

Our quantitative analysis from Fig.~\ref{timescaleVStemp} is further evidence for the model presented above. As expected from a thermally-activated oxygen depletion process, the loss increase time scale is observed to be exponential in the inverse of temperature $T$, i.e. $\tau_{th} \propto \exp(U/k_BT)$, where $U$ is some activation energy barrier, and $k_B$ is Boltzmann's constant. We find that the thermal activation energy barrier $U = 0.6(1)$~eV we obtained by fitting data in Fig.~\ref{timescaleVStemp} is in quantitative agreement with the difference in binding energy of $\sim 0.7$~eV, as measured by XPS, between Ta$_{2}$O$_{5}$ and its closest suboxide \cite{Demiryont1985}.

\subsection{Absence of Deposition Processes}

An alternative explanation for the observed loss process is spurious impurity deposition on the mirror surface, accelerated by higher temperatures, and reversed by the binding of impurities to ambient oxygen during the recovery process. However this type of model is inconsistent with our experimental setup and our observations. The chambers are maintained at vacuum levels making a deposition process unlikely. Most importantly, the observation that the loss increase rate depends on the thickness of the SiO$_2$ film making up the surface layer (Fig.~\ref{fig:422377SiTaCompFdecay}) refutes the possibility of a deposition process which is necessarily independent of film thicknesses.

\subsection{Absence of Measurement Light Effects on Loss Increase}

We note here that the loss increase we repeatedly observe (Section 3) is very unlikely to be significantly affected by our measurement light. The reasoning for this is two-fold. 

Under measurement conditions, the cavities are excited using a $\sim$100~$\mu$W source of resonant light, resulting in at most 10~kW/cm$^{2}$ of peak intra-cavity intensity, which is less than 0.1\% of the $4\times10^4$~kW/cm$^{2}$ damage threshold intensity reported for SiO$_2$:Ta$_{2}$O$_{5}$ dielectric stacks \cite{Zhao2003a}. This makes nonlinear light-induced losses a very unlikely mechanism to explain our observations. 

A loss mechanism that is linear in the integrated light intensity could remain. However, comparing light exposure of the cavity from chamber I to light exposure of the cavity from chamber II excludes a loss mechanism that is linear in the integrated light intensity. In chamber I (and chamber III), the cumulative cavity illumination time, when intra-cavity light is present for the purposes of obtaining free-decay traces, is a small fraction of the total experimental time for the loss increase data. Laser light is directed at the cavities only when measurements are made, while the measurement time sufficient to acquire statistics for each loss value (10-30 minutes on average) is much smaller than the time interval between measurements (ranging from a few hours to weeks). The cumulative illumination time of the mirrors per loss value measurement ($\leq$ 1 minute), at an intensity sufficient for measuring the free-decay traces, also represents a small fraction of the time for each loss value measurement (10-30 minutes). The integrated light intensity for data sets taken in chamber I (see Fig.~\ref{finesseVStemp}, data taken at 21$^{\circ}$C, 50$^{\circ}$C, 75$^{\circ}$C, 100$^{\circ}$C, and 150$^{\circ}$C) is therefore approximately $10$~kW/cm$^{2} \times 10$~min $=6\times 10^{6}$~J/cm$^2$. This is to be compared with the integrated light intensity exposure in chamber II, where to perform experiments unrelated to this paper, the cavity is stabilized relative to a 1~mW source of resonant light, resulting in $\sim$100~kW/cm$^{2}$ of continuous intra-cavity intensity. This high light intensity is present inside the cavity for hours at a time, which is repeated for hundreds of days over the course of observing the loss increase. This represents an integrated light intensity of approximately $100$~kW/cm$^{2} \times 1000$~hours $ \approx 4 \times 10^{11}$~J/cm$^2$. Notwithstanding this five orders of magnitude larger integrated light intensity, the loss increase rate measured for Coating I-1 in chamber II agrees qualitatively with what would be expected from extrapolating the loss increase rates measured for the same coating in chamber I (see Fig.~\ref{timescaleVStemp}). This excludes a loss increase process based on integrated light intensity.

We therefore conclude that our measurement light has no significant effect on the increase of mirror loss.

\section{Conclusion}

We conclude that the additional losses observed in mirror coatings placed under high vacuum are a result of a thermally-activated depletion of oxygen from the mirror's surface Ta$_2$O$_5$ layer. This process increases the concentration of absorbing TaO$_x$-suboxides, leading to a time-evolution of the loss factor that likely follows an Arrhenius process. This degradation process is strongly accelerated by temperature, with a rate that likely follows an exponential dependence on $1/T$. The loss can be reversed, in full or in part (depending on operating temperature), by filling the vacuum chamber to an atmospheric pressure of oxygen. The recovery from mirror loss can be strongly accelerated and enhanced by the presence of UV light. Most importantly for future systems, the loss process can be altogether prevented by passivating the Ta$_2$O$_5$ surface layer with a thin layer of SiO$_2$, on the order of 10~nm, or by ensuring that the surface layer of the dielectric stack is SiO$_2$.

\section*{Acknowledgments}

We thank the NSF-funded Center for Ultracold Atoms and the MQCO Program with funding from IARPA. D.G., A.B., and B.B. acknowledge support from the NSERC postgraduate scholarship program. M.G. acknowledges support from the NSF iQuISE IGERT program. Part of the results were obtained using facilities provided by the MIT NanoStructures Laboratory and MIT Microsystems Technology Laboratories.

\end{document}